\documentclass[nonblindrev]{informs3} 

\usepackage{balance}
\usepackage{amsmath}
\usepackage[tight,footnotesize]{subfigure}
\usepackage{graphicx}
\usepackage{bm}
\usepackage{algorithm}
\usepackage[noend]{algorithmic}
\usepackage{multirow}
\usepackage{caption}
\usepackage{amsfonts}

\OneAndAHalfSpacedXII 

\usepackage{natbib}
 \bibpunct[, ]{(}{)}{,}{a}{}{,}%
 \def\bibfont{\small}%
\TheoremsNumberedThrough     

\EquationsNumberedThrough    

\MANUSCRIPTNO{2015} 

\begin{document}


\RUNAUTHOR{Yuan and Tang}

\RUNTITLE{Worst-Case Adaptive Submodular Cover}

\TITLE{Worst-Case Adaptive Submodular Cover}

\ARTICLEAUTHORS{%
\AUTHOR{Jing Yuan}
\AFF{Department of Computer Science and Engineering, The University of North Texas}
\AUTHOR{Shaojie Tang}
\AFF{Naveen Jindal School of Management, The University of Texas at Dallas}
} 

\ABSTRACT{In this paper, we study the adaptive submodular cover problem under the worst-case setting. This problem generalizes many previously studied problems, namely,  the pool-based active learning and the stochastic submodular set cover.  The input of our problem is a set of items (e.g., medical tests) and each item has a random state (e.g., the outcome of a medical test), whose realization is initially unknown. One must select an item at a fixed cost in order to observe its realization. There is an utility function which maps a subset of items and their states to a non-negative real number. We aim to sequentially select a group of items to achieve a ``target value'' while minimizing the maximum cost across realizations (a.k.a. worst-case cost). To facilitate our study, we assume that the utility function is \emph{worst-case submodular}, a property that is commonly found in many machine learning applications. With this assumption, we develop a tight $(\log (Q/\eta)+1)$-approximation policy, where $Q$ is the ``target value'' and $\eta$ is the smallest difference between $Q$ and any achievable utility value $\hat{Q}<Q$. We also study a worst-case maximum-coverage problem, a dual problem of the minimum-cost-cover problem, whose goal is to select a group of items to maximize its worst-case utility subject to a budget constraint. To solve this problem, we develop a  $(1-1/e)/2$-approximation solution.}


\maketitle
\section{Introduction}
In this paper, we study a fundamental problem of minimum cost adaptive submodular cover under the worst-case setting. The problem can be formulated as follows: Given a set of items, each item has a state whose value is random and unknown initially, one must select an item at a fixed cost before observing its realized state. In addition, there is an utility function that depends on both the set of selected items and their realized states. Our goal is to sequentially select a group of items based on feedback,  in the form of the realized states of the selected items, to achieve a threshold function value at the minimum worst-case cost. Here the worst-case cost of a solution (a.k.a. policy)  refers to the maximum incurred cost  across realizations.  This formulation captures many  real-world applications, namely, active learning, viral marketing and sensor placement \citep{golovin2011adaptive}. As a motivating example, consider the application of medical diagnosis. Here each item represents a medical test and the state of an item refers to the outcome from corresponding medical test. Clearly, we can not observe the outcome of a test before performing that test. We define the utility function, in terms of a set of performed tests and their outcomes, as the number of false hypotheses  (e.g., diseases) ruled out by these tests. Suppose each test has a fixed cost, we aim to perform a sequence of tests (based on the outcomes from past tests) to eliminate \emph{all} false hypotheses at the minimum worst-case cost.

The minimum-cost adaptive submodular cover problem has received significant attention in the literature, however, most of the existing studies focus on minimizing the \emph{expected} cost of a policy \citep{golovin2011adaptive,esfandiari2021adaptivity,cui2022minimum}. In particular, they often assume that there is a known prior distribution over realizations, hence, they aim to find a policy that achieves  a threshold function value while minimizing the expected cost with respect to this distribution. In contrast, we focus on minimizing the worst-case cost of a policy, this is because in many real-world applications, it is often difficult or impossible to get an accurate prediction of how likely certain outcomes are.  Moreover, in many time-critical diagnostic applications, such as emergency response, one must rapidly identify a cause through a series of queries. In these applications, violation of a cost-constraint (such as time-constraint) may lead to fatal consequences; therefore, it is preferable to have a policy that has a small worst-case cost.

To solve this problem, we first introduce the concept of \emph{worst-case submodularity} \cite{doi:10.1287/ijoc.2022.1239},  extending the classic notation of \emph{submodularity} from sets to policies. We say a function is worst-case submodular if the worst-case marginal utility of an item satisfies the diminishing returns property (Definition \ref{def:11}). This property is prevalent across a diverse range of applications such as the pool-based active learning and the stochastic submodular set cover. Our main contribution is to develop a best possible $(\log (Q/\eta)+1)$-approximation policy for the worst-case adaptive submodular cover problem, where $Q$ is the ``target value'' and $\eta$ is the smallest difference between $Q$ and any achievable  utility value $\hat{Q}<Q$. In addition, we study a worst-case maximum-coverage problem, whose goal is to sequentially select a group of items to maximize its worst-case utility subject to a budget constraint. We develop a $(1-1/e)/2$-approximation solution for this problem.

\paragraph{Additional related works.} There is some work on minimizing the worst-case cost in active learning; see e.g., \citep{cicalese2017decision,moshkov2010greedy}. Our results can be viewed as a generalization of their results because we can show that the utility function of pool-based active learning (or optimal decision tree design in general) is worst-case submodular. Recently, \cite{golovin2011adaptive} introduced the concept of \emph{adaptive submodularity}. Similar to our notation of worst-case submodularity, adaptive submodularity is another way of extending submodularity from sets to policies. However, their property depends on the prior distribution of realizations, whereas there is no such dependence in defining worst-case submodularity. More importantly, our proposed notation allows for better approximation bounds in many real-world applications. In particular, \cite{golovin2011adaptive} developed a $(1+\log \frac{Q}{\eta p_{\min}})$-approximation policy for the minimum-cost coverage problem under the worst-case setting if the utility function is adaptive submodular, where $p_{\min}$ is the minimum probability of any realization. In contrast, our policy achieves the $(1+\log (Q/\eta))$-approximation bound;  $1/p_{\min}$ can  be exponentially larger than $Q$. It is also worth noting that one must know the prior distribution over realizations in order to implement \cite{golovin2011adaptive}'s policy whereas ours does not need such information. Finally, \citep{guillory2010interactive,guillory2011simultaneous} studied the simultaneous learning and covering problem, whereas we focus on the covering problem. The problem of constrained adaptive submodular maximization has been widely researched in the literature. Most of the existing studies center on maximizing the average-case utility \citep{golovin2011adaptive,tang2021beyond,tang2021beyond2,tang2021adaptive,tang2022partial,tang2021non,tang2022optimal,tang2022group} whereas our focus is on maximizing worst-case utility. The concept of worst-case submodularity was recently introduced by \cite{doi:10.1287/ijoc.2022.1239} where they studied the worst-case submodular maximization problem subject to matroid constraints, we examine the same problem subject to a different constraint, such as budget constraints, instead.

\section{Preliminaries}
In the rest of this paper, we use $[m]$ as shorthand notation for the set $\{1, 2, \cdots, m\}$.

\subsection{Items and States.} The input is a set  $E$ consisting of $n$ items.  Each item $e \in E$ is in an undetermined state from a set $O$ of possible states. We use a function $\phi : E \rightarrow O$, called a \emph{realization}, to represent the item states, where function $\phi$ maps each item in the ground set $E$ to a state in $O$. Therefore, we can say that $\phi(e)$ represents the state of $e$ under the realization $\phi$. In the example of diagnosis, each item $e$ represents a medical test and $\phi(e)$ is the outcome of $e$.  We use $\Phi$ to represent a randomly determined realization. One must select one item in order to uncover its realized state.  We assume that selecting an item $e$ incurs a fixed cost $c(e)$. For convenience, let $c(S)=\sum_{e\in S} c(e)$.

For any subset of items  $S\subseteq E$, we use the notation $\psi: S\rightarrow O$ to represent a \emph{partial realization} of $S$. Let $\mathrm{dom}(\psi)=S$ denote the \emph{domain} of $\psi$.
Consider a realization $\phi$ and a partial  realization $\psi$,  we say that $\phi$ is consistent with $\psi$ (denoted as $\phi \sim \psi$) if $\phi$ and $\psi$ are equal everywhere in $\mathrm{dom}(\psi)$.  We say that a partial realization $\psi$  is a \emph{subrealization} of another partial realization $\psi'$ (denoted  as $\psi \subseteq \psi'$) if the two realizations are identical in the domain of $\psi$ (i.e., $\mathrm{dom}(\psi)$) and  $\mathrm{dom}(\psi)$ is a subset of $\mathrm{dom}(\psi')$.


\subsection{Policy and Worst-Case Submodularity}
\label{sec:3.2}
Under the adaptive setting, we aim to find an adaptive solution which selects items sequentially and adaptively,  with each selection being based on the previously obtained feedback. Formally, any adaptive solution can be represented as a policy $\pi$ that maps the current observation to the next item to be selected: $\pi: 2^{E}\times O^E \rightarrow E$.  For example, suppose we observe a partial realization $\cup_{e\in S}\{ (e, \Phi(e))\}$ after selecting a set $S$ of items and assume  $\pi(\cup_{e\in S} \{(e, \Phi(e))\})= w$. Then $\pi$ selects $w$ as the next item. It is certainly possible to define a randomized policy by mapping the current observation to some distribution of items. However, because every randomized policy can be considered as a distribution of a set of deterministic policies, we focus on deterministic policies without loss of generality. 



There is a utility function $f: 2^{E}\times 2^{O} \rightarrow \mathbb{R}_{\geq0}$ which maps a subset of items and their states to a non-negative real number. Let $E(\pi, \phi)$ denote the subset of items selected by the policy $\pi$ under the realization $\phi$. 
Let $U^+$ denote the set of all realizations that have a \emph{positive} probability of occurring. The worst-case  utility, $f_{wc}(\pi)$, of a policy $\pi$ is defined as the minimum utility that can be achieved by $\pi$ over all possible realizations, it can be written as
\begin{eqnarray}
f_{wc}(\pi)=\min_{\phi\in U^+}f(E(\pi, \phi), \phi).~\nonumber
\end{eqnarray}

For ease of presentation, we extend the definition of $f$ by letting $f(\psi) = \mathbb{E}_{\Phi}[f(\mathrm{dom}(\psi), \Phi)\mid \Phi \sim \psi]$ denote the expected utility of $\mathrm{dom}(\psi)$ conditioned on the partial realization $\psi$.  We now present the concept of worst-case marginal utility $\Delta_{wc}(e \mid \psi)$  of item $e$  when added to a partial realization $\psi$. Let $p(\phi\mid \psi) =\Pr[\Phi=\phi\mid \Phi\sim \psi ]$ denote the conditional distribution over realizations conditioned on  a partial realization $\psi$. Define
\begin{eqnarray}
\label{def:wc}
\Delta_{wc}(e \mid \psi)=\min_{o \in O(e, \psi)}\{f(\psi\cup\{(e, o)\})-f(\psi)\},~\nonumber
\end{eqnarray} where $ O(e, \psi)=\{o\in O\mid \exists \phi: p(\phi\mid\psi)>0, \phi(e)=o\}$ denotes the set of \emph{possible} states that $e$ can take on, given the partial realization $\psi$.

Now we are ready to introduce the notations of worst-case submodularity and worst-case monotonicity \cite{doi:10.1287/ijoc.2022.1239}.
\begin{definition}\emph{[Worst-case Submodularity and Worst-case Monotonicity]} A function   $f$  is  worst-case submodular if
\begin{eqnarray}\label{def:332}
\Delta_{wc}(e\mid \psi) \geq \Delta_{wc}(e\mid \psi')
\end{eqnarray} for any two partial realizations $\psi$ and $\psi'$ such that $\psi\subseteq \psi'$ and for any item   $e\in E\setminus \mathrm{dom}(\psi')$.
A function $f$ is worst-case monotone if for every partial realization $\psi$ and any $e\in E\setminus \mathrm{dom}(\psi)$, $\Delta_{wc}(e\mid \psi) \geq 0$.
\label{def:11}
\end{definition}

Lastly, we introduce the concept of \emph{minimal dependency} \citep{cuong2014near}, which states that the utility of any collection of items is only dependent on the state of the items within that group.
\begin{definition}\emph{[Minimal Dependency]} We say a function $f$ is minimal dependent with respect to $p(\phi)$ if for any partial realization $\psi$ and any realization $\phi$ such that $\phi\sim \psi$, we have $f(\mathrm{dom}(\psi)) = f(\mathrm{dom}(\psi), \phi)$. 
\end{definition}

The properties of worst-case submodular, worst-case monotone and minimal dependent can be observed in a wide range of applications, such as pool-based active learning, stochastic submodular set cover, and adaptive influence maximization. Therefore, all results derived in this paper are applicable to these types of applications.
\section{Problem Formulation}
 Given a policy $\pi$, let $c_{wc}(\pi)$ denote the worst-case cost of $\pi$, formally, $c_{wc}(\pi)=\max_{\phi\in U^+} c(E(\pi,\phi))$. We assume there is a ``target value'' $Q$ such that $f(E, \phi)=Q$ for all $\phi$. The worst-case adaptive submodular cover problem  is formally defined as follows:
\[\min_{\pi:  f_{wc}(\pi) \geq Q}c_{wc}(\pi).\]

For the case if $f(E, \phi)$ varies across $\phi$, we can define a new function $\hat{f}(S, \phi)=\min\{Q', f(S, \phi)\}$, where $Q'$ is some
threshold that is no larger than $\min_{\phi} f(E, \phi)$; that is, $Q'$ is achievable under all realizations. Fortunately, this variation does not add additional difficulty to our problem because Lemma \ref{lem:aaa} shows that if $f$ is worst-case monotone, worst-case submodular and minimal dependent with respect to $p(\phi)$, then $\hat{f}$ is also  worst-case monotone, worst-case submodular and minimal dependent with respect to $p(\phi)$, indicating that our results still hold if we replace the original utility function $f$  and the ``target value'' $Q$ with $\hat{f}$ and $Q'$, respectively.

\begin{lemma}
\label{lem:aaa}
Let $\hat{f}(S, \phi)=\min\{Q', f(S, \phi)\}$ for some constant $Q'$. If $f$ is worst-case monotone, worst-case submodular and minimal dependent with respect to $p(\phi)$, then $\hat{f}$ is also  worst-case monotone, worst-case submodular and  minimal dependent with respect to $p(\phi)$.
\end{lemma}
\emph{Proof:}  It is trivial to show that if  $f$ is worst-case monotone and minimal dependent, then $\hat{f}$ is also  worst-case monotone and  minimal dependent. We next focus on proving that if $f$ is  worst-case submodular, then $\hat{f}$ is also worst-case submodular. We start by  presenting a useful technical lemma in Lemma \ref{lem:end}. Its proof is provided in appendix.

\begin{lemma}
\label{lem:end}
Consider any five constants $c_1, c_2, c_3, c_4$ and $x$ such that $c_1\geq c_2$ and $c_3 \geq c_4$, $c_1-c_2\geq c_3-c_4$ and $c_2 \leq c_4$, we have $\min\{c_1, x\}-\min\{c_2, x\}\geq\min\{c_3, x\}-\min\{c_4, x\}$.
\end{lemma}

Consider any two partial realizations $\psi$ and $\psi'$ such that $\psi\subseteq \psi'$ and any   $e\in E\setminus \mathrm{dom}(\psi')$,
{\small\begin{eqnarray*}
&&\min_{o \in O(e, \psi)}\left\{\hat{f}(\psi\cup\{(e, o)\})-\hat{f}(\psi)\right\}\\
&&\quad\quad\quad\quad\quad\quad-\min_{o \in O(e, \psi')}\left\{\hat{f}(\psi'\cup\{(e, o)\})-\hat{f}(\psi')\right\}\\
&&= \min_{o \in O(e, \psi)}\left\{\min\{Q', f(\psi\cup\{(e, o)\})\}-\min\{Q', f(\psi)\}\right\}\\
&&\quad-\min_{o \in O(e, \psi')}\left\{\min\{Q', f(\psi'\cup\{(e, o)\})\}-\min\{Q', f(\psi')\}\right\}\\
&&= \left(\min\{Q', \min_{o \in O(e, \psi)} f(\psi\cup\{(e, o)\})\}-\min\{Q', f(\psi)\}\right)~\nonumber\\
&&-\left(\min\{Q', \min_{o \in O(e, \psi')}f(\psi'\cup\{(e, o)\})\}-\min\{Q', f(\psi')\}\right).\label{eq:1}
\end{eqnarray*}}

To prove the worst-case submodularity of $\hat{f}$, it suffices to show that
{\small\begin{eqnarray} &&\left(\min\{Q', \min_{o \in O(e, \psi)} f(\psi\cup\{(e, o)\})\}-\min\{Q', f(\psi)\}\right) \label{eq:rain}\\
&&\hspace*{-0.7cm}-\left(\min\{Q', \min_{o \in O(e, \psi')}f(\psi'\cup\{(e, o)\})\}-\min\{Q', f(\psi')\}\right)\geq0.~\nonumber
\end{eqnarray}}

Let $c_1=\min_{o \in O(e, \psi)} f(\psi\cup\{(e, o)\})$, $c_2=f(\psi)$, $c_3=\min_{o \in O(e, \psi')}f(\psi'\cup\{(e, o)\})$ and $c_4= f(\psi')$, we have  $c_1\geq c_2$ (by worst-case monotonicity), $c_3 \geq c_4$ (by worst-case monotonicity), $c_1-c_2\geq c_3-c_4$ (by worst-case submodularity) and $c_2 \leq c_4$ (by worst-case monotonicity). Hence, apply Lemma \ref{lem:end} with these parameters gives inequality (\ref{eq:rain}).
$\Box$

\section{Algorithm Design and Analysis}
\begin{algorithm}[hptb]
\caption{Worst-Case Density-Greedy Policy $\pi^g$}
\label{alg:LPP3}
\begin{algorithmic}[1]
\STATE $t=1; \psi_0=\emptyset$.
\WHILE {$f(\psi_{t})< Q$}
\STATE select $e_t \in \argmax_{e\in E} \frac{\Delta_{wc}(e\mid \psi_{t-1})}{c(e)}$;
\STATE observe $\Phi(e_t)$ and update $\psi_{t}\leftarrow \psi_{t-1}\cup\{(e_t, \Phi(e_t))\}$;
\STATE $t\leftarrow t+1$;
\ENDWHILE
\end{algorithmic}
\end{algorithm}

We first introduce a Worst-Case Density-Greedy Policy (labeled as $\pi^g$) for the worst-case adaptive submodular cover problem. In each step $t$ of $\pi^g$, it selects an item $e_t$ that maximizes the worst-case ``benefit-to-cost'' ratio on top of the current observation, i.e.,
\begin{eqnarray}
e_t \in \argmax_{e\in E} \frac{\Delta_{wc}(e\mid \psi_{t-1})}{c(e)},
\end{eqnarray}
where $\psi_{t-1}$ denotes the partial realization observed at step $t$. Then it updates the observation using $\psi_{t}\leftarrow \psi_{t-1}\cup\{(e_t, \Phi(e_t))\}$. We follow this density-greedy rule to select items recursively until the utility of selected items achieves the quality threshold $Q$, i.e., $f(\psi_{t})\geq Q$. With the assumption that $f$ is minimal dependent, it is easy to verify that $f_{wc}(\pi^g)\geq Q$. A detailed description of $\pi^g$ is listed in Algorithm \ref{alg:LPP3}.


We conduct our analysis based on the concept of  \emph{virtual slot}, which was originally proposed in \cite{golovin2011adaptive}. Assume after a policy $\pi$ selects an item $e$, it starts to ``run'' $e$, and terminates after $c(e)$ virtual slots. It is worth noting that virtual slot is only defined for analytical purposes and does not consume actual time. Based on this notation, we introduce the level-$l$-truncation  $\pi_l$  of a policy $\pi$  over virtual time as follows.

 \begin{definition}[Level-$l$-truncation of $\pi$ over virtual time] Run $\pi$ for $l$ virtual slots, and for every item $e\in E$, if  $e$ has been running for $\gamma$ virtual slots, selecting $e$ independently with probability $\gamma/c(e)$. 
 \end{definition}

For example, assume a policy $\pi$ selects three items $e_1$, $e_2$, $e_3$ in the end with $c(e_1)=2$, $c(e_2)=2$ and $c(e_3)=3$. Then its level-$5$-truncation  $\pi_5$ selects $e_1$ and $e_2$ deterministically, and selects $e_3$ with probability $1/3$; its level-$3$-truncation  $\pi_3$ selects $e_1$  deterministically, and selects $e_2$ with probability $1/2$.

 Given a realization $\phi$ and a policy $\pi$, for any $l\in \mathbb{Z}^+$, let $t[l, \phi, \pi]$ denote the number of items that have a positive probability of being selected by $\pi_l$ conditioned on $\phi$. For convenience, we use $t[l]$ to denote $t[l, \phi, \pi]$ if it is clear from the context. In the previous example, we have  $t[3]=2$ because both $e_1$ and $e_2$ have a positive probability of being selected by $\pi_{3}$; we have $t[5]=3$ because all three items have a positive probability of being selected by $\pi_{5}$.

We denote with $h(\pi_l\mid \phi)$ the expected utility of $\pi_l$ conditioned on a realization $\phi$. Assume $\psi_0=\emptyset$.  With the above notations and the definition of $\pi_l$, $h(\pi_l\mid \phi)$ is formally defined as follows:
{\small\begin{eqnarray}
&&h(\pi_l\mid \phi)=  f(\psi_{t[l]-1})+ \label{eq:cold}\\
&&\frac{\min\{l-c(\mathrm{dom}(\psi_{t[l]-1})), c(e_{t[l]})\}}{c(e_{t[l]})}\left( f(\psi_{t[l]})-f(\psi_{t[l]-1})\right), ~\nonumber
 \end{eqnarray}}
where $f(\psi_{t[l]-1})$ is the utility of the first $t[l]-1$ items (i.e., $\mathrm{dom}(\psi_{t[l]-1})$) that are selected by $\pi_l$ deterministically, $\frac{\min\{l-c(\mathrm{dom}(\psi_{t[l]-1})), c(e_{t[l]})\}}{c(e_{t[l]})}$ is the selection probability of the $t[l]$-th item, and
$f(\psi_{t[l]})-f(\psi_{t[l]-1})$ is the utility of the $t[l]$-th item.

Before providing the main theorem, we first present two technical results.
\begin{lemma}
\label{lem:museum}
Given any realization $\phi$ and a policy $\pi$, for any $l\leq c(E(\pi, \phi))$, we have
\begin{eqnarray}
h(\pi_l\mid \phi) - h(\pi_{l-1}\mid \phi)= \frac{1}{c(e_{t[l]})}\left( f(\psi_{t[l]})-f(\psi_{t[l]-1})\right).~\nonumber
\end{eqnarray}
\end{lemma}
\emph{Proof:} Observe that if $l\leq c(E(\pi, \phi))$, then
\begin{eqnarray}
c(\mathrm{dom}(\psi_{t[l]-1}))+c(e_{t[l]})=c(\mathrm{dom}(\psi_{t[l]})) \geq l.
\end{eqnarray} Hence, $\min\{l-c(\mathrm{dom}(\psi_{t[l]-1})), c(e_{t[l]})\}=l-c(\mathrm{dom}(\psi_{t[l]-1}))$. It follows that (\ref{eq:cold}) can be simplified to
{\small\begin{eqnarray}
&&h(\pi_l\mid \phi)= \label{eq:cold1} \\
&&f(\psi_{t[l]-1})+ \frac{l-c(\mathrm{dom}(\psi_{t[l]-1}))}{c(e_{t[l]})}\left( f(\psi_{t[l]})-f(\psi_{t[l]-1})\right). ~\nonumber
 \end{eqnarray}}
To prove this lemma, we consider two cases:

\textbf{Case 1:} We first consider the case when $t[l]=t[l-1]$. Observe that for any $l\leq c(E(\pi, \phi))$, it holds that
{\small\begin{eqnarray*}
&&h(\pi_l\mid \phi) - h(\pi_{l-1}\mid \phi)= \\
&& \left(f(\psi_{t[l]-1}) + \frac{l-c(\mathrm{dom}(\psi_{t[l]-1}))}{c(e_{t[l]})}\left( f(\psi_{t[l]})-f(\psi_{t[l]-1})\right)\right)\\
&&- \Bigg(f(\psi_{t[l-1]-1}) + \\
&& \frac{l-1-c(\mathrm{dom}(\psi_{t[l-1]-1}))}{c(e_{t[l-1]})}\left( f(\psi_{t[l-1]})-f(\psi_{t[l-1]-1})\right)\Bigg)\\
&& = \frac{1}{c(e_{t[l]})}\left( f(\psi_{t[l]})-f(\psi_{t[l]-1})\right),
 \end{eqnarray*}}
where the first equality is from (\ref{eq:cold1}) and the second equality is from the assumption that  $t[l]=t[l-1]$.

\textbf{Case 2:} We next consider the case when $t[l]=t[l-1]+1$, that is, $l-1$ is the last virtual slot in round $t[l-1]$ and $l$ is the first virtual slot in round $t[l]$. In this case, we can rewrite $h(\pi_l\mid \phi)$ as
{\small \begin{eqnarray}
&&h(\pi_l\mid \phi)= f(\psi_{t[l]-1}) + \frac{1}{c(e_{t[l]})}\left( f(\psi_{t[l]})-f(\psi_{t[l]-1})\right)~\nonumber\\
&& = f(\psi_{t[l-1]}) + \frac{1}{c(e_{t[l]})}\left( f(\psi_{t[l]})-f(\psi_{t[l-1]})\right), \label{eq:kenzie}
 \end{eqnarray}}
 where the first equality is by (\ref{eq:cold1}) and the observation that $l$ is the first virtual slot in round $t[l]$ and the second equality is by the assumption that $t[l]=t[l-1]+1$. Meanwhile,
{\small \begin{eqnarray}
&&h(\pi_{l-1}\mid \phi)= ~\nonumber\\
&&f(\psi_{t[l-1]-1})+ ~\nonumber\\
&&\quad\quad\quad\quad\frac{l-c(\mathrm{dom}(\psi_{t[l-1]-1}))}{c(e_{t[l-1]})}\left( f(\psi_{t[l-1]})-f(\psi_{t[l-1]-1})\right)~\nonumber\\
&& = f(\psi_{t[l-1]-1}) + \frac{c(e_{t[l-1]})}{c(e_{t[l-1]})}\left( f(\psi_{t[l-1]})-f(\psi_{t[l-1]-1})\right)~\nonumber\\
&&= f(\psi_{t[l-1]}), \label{eq:ivory}
 \end{eqnarray}}
where the second equality is because $l-1$ is the last virtual slot in round $t[l-1]$, indicating that $l-c(\mathrm{dom}(\psi_{t[l-1]-1}))=c(e_{t[l-1]})$. Equalities (\ref{eq:kenzie}) and (\ref{eq:ivory}) together imply that $h(\pi_l\mid \phi) - h(\pi_{l-1}\mid \phi)= \frac{1}{c(e_{t[l]})}\left( f(\psi_{t[l]})-f(\psi_{t[l]-1})\right)$.
$\Box$

Throughout the rest of this paper, let $c^*=c_{wc}(\pi^*)$ denote the worst-case cost of the optimal solution $\pi^*$.
\begin{theorem}
\label{thm:2}
 If the utility function $f$  is worst-case monotone and worst-case submodular with respect to $p(\phi)$ , then for any $L\in \mathbb{Z}^+$ and any realization $\phi$, it holds that
 \begin{eqnarray}
\label{eq:last111}
h(\pi^g_L\mid \phi)> (1-e^{-L/c^*}) Q,
\end{eqnarray} where $\pi^g_L$ is the level-$L$-truncation of $\pi^g$.
\end{theorem}
\emph{Proof:} We first recall some notations. For each $t\in[n]$, let $\psi_t$ represent the partial realization of the first $t$ items picked by $\pi^g$ conditioned on $\phi$. We use $t[l]$ to denote the number of items that have a positive probability of being selected by $\pi^g_l$ conditioned on $\phi$. Hence, $\psi_{t[l]-1}$ represents the partial realization of the first $t[l]-1$  items selected by $\pi^g_l$ conditioned on $\phi$.

The case when $L> c(E(\pi^g, \phi))$ is trivial. If $L> c(E(\pi^g, \phi))$, then $t[L]=|E(\pi^g, \phi)|$ by the definition of $t[L]$. It follows that $h(\pi^g_L\mid \phi)= f(\psi_{t[L]})=Q$, where the first equality is by the definition of $h$ and the assumption that $L> c(E(\pi^g, \phi))$;  the second equality is by the observation that $\pi^g$  achieves the target value $Q$ after selecting all $t[L]$ items. Next we focus on the case when $L\leq c(E(\pi^g, \phi))$.

 Given any $\psi_{t[l]-1}$,  we create a realization $\phi^*$ in the following way. First, we make sure that $\phi^*$ is consistent with $\psi_{t[l]-1}$ by defining $\phi^*(e)=\phi(e)$ for each $e\in \mathrm{dom}(\psi_{t[l]-1})$. For the rest of the items, we decide their states in $\phi^*$ incrementally by simulating the execution of the optimal policy $\pi^*$ conditioned on $\psi_{t[l]-1}$. Let $\psi^*_i$ denote the partial realization after running $\pi^*$ for $i$ rounds. Starting with $i=1$ and let $\psi^*_0=\emptyset$, in each subsequent round $i$, assume  $\pi^*$ selects $e^*_i$ as the $i$-th item after observing $\psi^*_{i-1}$, we define the state of $e^*_i$ in $\phi^*$ as follows:
\begin{eqnarray*}
&&\phi^*(e^*_i) = \argmin_{o\in O(e^*_i, \psi_{t[l]-1}\cup \psi^*_{i-1})}f(\psi_{t[l]-1}\cup \psi^*_{i-1}\cup \{(e^*_i, o)\}).
 \end{eqnarray*}
The observation $\psi^*_{i}$ is updated by adding new information from $(e^*_i, \phi^*(e^*_i))$ and the previous observation $\psi^*_{i-1}$, and then $\pi^*$ proceeds to the next round. This continues until $\pi^*$ terminates, at which point the states of all items selected by $\pi^*$ have been determined. The intuition behind creating such $\phi^*$ is that in each round $i$, we pick a state that is the least favorable for $e^*_i$, in order to decrease the marginal utility of adding $e^*_i$ to the partial realization $\psi_{t[l]-1}\cup \psi^*_{i-1}$ as much as possible. Without loss of generality,  it can be assumed that $\pi^*$ ends up choosing $k$ items. It is possible that there are multiple realizations that fit this description, one of them is  arbitrarily chosen as $\phi^*$; in particular, $\phi^*$ could be any realization that is consistent with $\psi_{t[l]-1}\cup \psi^*_k$.

To prove this theorem, it suffices to show that for all $l\in[L]$, it holds that

\begin{eqnarray}h(\pi^g_l\mid \phi) - h(\pi^g_{l-1}\mid \phi)\geq  \frac{ Q-h(\pi^g_{l-1}\mid \phi)}{c^*}. \label{eq:f}
\end{eqnarray}

This is because by induction on $l$, we have that for any $L\in \mathbb{Z}^+$,
\begin{eqnarray}
h(\pi^g_L\mid \phi) &=& \sum_{l\in [L]} \left( h(\pi^g_l\mid \phi) - h(\pi^g_{l-1}\mid \phi) \right)~\nonumber\\
&>& (1-e^{-L/c^*}) Q.\label{eq:last}
\end{eqnarray}

We will concentrate on demonstrating (\ref{eq:f}) for the remainder of the proof. Let $e_{t[l]}$ denote the $t[l]$-th item selected by $\pi^g$ conditioned on $\phi$, the following chains proves (\ref{eq:f}):
\begin{eqnarray}
&&h(\pi_l\mid \phi) - h(\pi_{l-1}\mid \phi)=   \frac{1}{c(e_{t[l]})}\left( f(\psi_{t[l]})-f(\psi_{t[l]-1})\right)~\nonumber\\
&&= \frac{1}{c(e_{t[l]})}\left( f(\psi_{t[l]-1}\cup \{(e_{t[l]}, \phi(e_{t[l]}))\})-f(\psi_{t[l]-1})\right)~\nonumber\\
&&\geq \min_{o\in O(e_{t[l]}, \psi_{t[l]-1})}\frac{\left(f( \psi_{t[l]-1}\cup\{(e_{t[l]}, o)\})-f(\psi_{t[l]-1})\right)}{c(e_{t[l]})}~\nonumber\\
&&= \max_{e\in E}\frac{1}{c(e)}\Delta_{wc}(e\mid \psi_{t[l]-1})\geq \max_{i\in[k]}\frac{\Delta_{wc}(e^*_i\mid \psi_{t[l]-1})}{c(e^*_i)} ~\nonumber\\
&&\geq \frac{\sum_{i\in [k]}\Delta_{wc}(e^*_i\mid \psi_{t[l]-1})}{\sum_{i\in [k]}c(e^*_i)}~\nonumber\\
&&\geq \frac{\sum_{i\in [k]}\Delta_{wc}(e^*_i\mid \psi_{t[l]-1})}{c^*}~\nonumber\\
&&\geq \frac{\sum_{i\in [k]}\Delta_{wc}(e^*_i\mid \psi_{t[l]-1}\cup\psi^*_{i-1})}{c^*}~\nonumber\\
&&=\frac{ f(\psi_{t[l]-1}\cup\psi^*_k)-f(\psi_{t[l]-1})}{c^*}\geq \frac{ f(\psi^*_k)-f(\psi_{t[l]-1})}{c^*} ~\nonumber\\
&&= \frac{ Q-f(\psi_{t[l]-1})}{c^*}\geq  \frac{ Q- h(\pi_{l-1}\mid \phi)}{c^*}~\nonumber.
\end{eqnarray}
The first equality is from the assumption that $L\leq c(E(\pi^g, \phi))$ and Lemma \ref{lem:museum}, the first inequality is due to $\phi(e_{t[l]})\in O(e_{t[l]}, \psi_{t[l]-1})$, the fourth inequality is by the definition of $c^*$, the fifth inequality is due to $f$ being worst-case submodular, the sixth inequality is due to $f$ being worst-case monotone, the last equality is because $\pi^*$ is a valid solution, indicating that $f(\psi^*_k)=Q$, and the last inequality is by the definition of $h(\pi_{l-1}\mid \phi)$ (Eq. (\ref{eq:cold})).
 $\Box$

We next present the main theorem of this section.

\begin{theorem}
\label{thm:21}
Suppose the utility function $f$  is worst-case monotone, worst-case submodular with respect to $p(\phi)$ and it satisfies the property of minimal dependency. Let $\eta$ be any value such that
$f(\psi) > Q-\eta$ implies $f(\psi) = Q$ for all partial realization $\psi$. Then $\pi^g$ is a feasible solution and $c_{wc}(\pi^g) \leq (\ln\frac{Q}{\eta}+1)c^*$.
\end{theorem}
\emph{Proof:} Let $\phi'$ denote the worst-case realization with respect to $\pi^g$, that is, $\phi'\in \arg\max_{\phi\in U^+} c(E(\pi^g, \phi))$.  Apply Theorem \ref{thm:2} with $L=c^*\ln(Q/\eta)$ and $\phi=\phi'$ to give
\begin{eqnarray}
\label{eq:last1}
 h(\pi^g_L\mid \phi') > (1-e^{-L/c^*}) Q = (1-\frac{\eta}{Q}) Q = Q-\eta.
\end{eqnarray}

Define $\pi^g_{L\rightarrow}$ as a policy that is identical to $\pi^g_L$ except that $\pi^g_{L\rightarrow}$ selects the $t[L]$-th item deterministically. Hence,
\begin{eqnarray}
\label{eq:last2}
h(\pi^g_{L\rightarrow} \mid \phi') \geq h(\pi^g_L\mid \phi') > Q-\eta,
\end{eqnarray}
where the first inequality is because $\pi^g_{L\rightarrow}$ selects the $t[L]$-th item deterministically while $\pi^g_L$ might select this item probabilistically, indicating that the utility of $\pi^g_{L\rightarrow}$ is no less than that of $\pi^g_L$; the second inequality is from (\ref{eq:last1}).

By the definition of $\eta$, we have
\begin{eqnarray}
\label{eq:last3}
h(\pi^g_{L\rightarrow} \mid \phi') = Q,
\end{eqnarray}
and moreover, $\pi^g$ must select $t[L]$ items.

Hence, the worst-case cost of $\pi^g$ is $c(E(\pi^g_{L\rightarrow}, \phi'))$. To prove this theorem, it suffices  to show that $c(E(\pi^g_{L\rightarrow}, \phi'))$ is upper bounded by $(\ln\frac{Q}{\eta}+1)c^*$.

To prove this bound, we first show that the cost of every item selected by $\pi^g$ is at most $c^*$. Consider any round $t$ of $\pi^g$, (\ref{eq:f}) and Lemma \ref{lem:museum} jointly imply that
\begin{eqnarray}
\label{eq:ian}
\frac{ f(\psi'_{t-1}\cup \{(e'_{t}, \phi'(e'_{t}))\})-f(\psi'_{t-1})}{c(e'_{t})}\geq  \frac{ Q-f(\psi'_{t-1})}{c^*},
\end{eqnarray}
where $\psi'_t$ represents the partial realization of the first $t$ items picked by $\pi^g$ conditioned on $\phi'$; $e'_{t}$ is the $t$-th item selected by $\pi^g$ conditioned on $\phi'$.

Because $f(\psi'_{t-1}\cup \{(e'_{t}, \phi'(e'_{t}))\})\leq Q$, we have $ f(\psi'_{t-1}\cup \{(e'_{t}, \phi'(e'_{t}))\})-f(\psi'_{t-1})\leq Q-f(\psi'_{t-1})$. This, together with (\ref{eq:ian}), implies that $c(e'_{t})\leq c^*$ for all $t$. This implies that the cost of  the $t[L]$-th item selected by $\pi^g$ is at most $c^*$, i.e., $c(e'_{t[L]})\leq c^*$. It follows that
\begin{eqnarray}
\label{eq:ian1}
c(E(\pi^g_{L\rightarrow}, \phi'))\leq L+c^* = c^*\ln(Q/\eta) +c^* = (\ln\frac{Q}{\eta}+1)c^*,
\end{eqnarray}
where the first equality is from the following observation: if $L> c(E(\pi^g, \phi'))$, then $c(E(\pi^g_{L\rightarrow}, \phi'))\leq c(E(\pi^g, \phi'))< L$; if $L\leq c(E(\pi^g, \phi'))$, then $c(E(\pi^g_{L\rightarrow}, \phi'))\leq c(E(\pi^g, \phi')) \leq L+c(e'_{t[L]})\leq L+c^*$. $\Box$

\textbf{Tightness of Our Results:} It is easy to verify that the classic \emph{deterministic} submodular cover problem \citep{wolsey1982analysis} is a special case of our problem. Given  that the best approximation ratio for the deterministic submodular cover problem is $\ln\frac{Q}{\eta}+1$, the guarantee provided in Theorem \ref{thm:21}
is the best possible.

\subsection{Pointwise submodularity is not sufficient}
\label{sec:doesnot}
A function $f$  is  called pointwise submodular if, $f(\cdot, \phi): 2^E\rightarrow \mathbb{R}_{\geq 0}$ is submodular for all realizations $\phi\in U^+$. This property can be found in numerous applications. Unfortunately, we next construct an example to show that  the ratio of  $c_{wc}(\pi^g)$ and $c^*$ could be arbitrarily large even if $f$ is pointwise submodular and $Q/\eta=1$. In other words, pointwise submodularity is not sufficient to guarantee the performance bound from Theorem \ref{thm:21}.

Consider a set of three items $E=\{e_1, e_2, e_3\}$ with cost $c(e_1)=\epsilon_a$ and $c(e_2)=c(e_3)=\epsilon_b$. 
There are two possible states $O=\{o_1, o_2\}$. Assume $U^+$ is composed of two possible realizations:
\[\phi_1=\{ (e_1, o_1), (e_2, o_1), (e_3, o_2)\}\] \[\phi_2=\{(e_1, o_1), (e_2, o_2), (e_3, o_1)\}\]

Therefore, $e_1$ has a deterministic state $o_1$, whereas $e_2$'s state is different from $e_3$'s state. We consider a modular utility function $f$ such that $f(S,\phi)=\sum_{e\in S} v_{e, \phi(e)}$, where $v_{e, \phi(e)}$ is the value of $e$ in state $\phi(e)$. We assume that $e_1$ has a deterministic value of $Q$; and $e_2$ (resp. $e_3$) has a value of $Q$ (resp. $0$) in state $o_1$ and a value of $0$ (resp. $Q$) in state $o_2$, that is, $v_{e_2, o_1}=v_{e_3, o_2}=Q$ and $v_{e_2, o_2}=v_{e_3, o_1}=0$. First, because $f$ is a linear function, it is also pointwise submodular. Moreover, it is easy to verify that $f$ is worst-case monotone and minimal dependent. Second, $\eta=Q$ in our example by the definition of $f$; hence,  $Q/\eta=1$.
 According to the design of $\pi^g$, it always selects $e_1$ because the worst-case ``benefit-to-cost'' ratio of $e_1$ (with respect to an empty set) is $Q/\epsilon_a> 0$, however, the worst-case ``benefit-to-cost'' ratios of $e_2$ and $e_3$ are both $0$.  By contrast, the optimal solution $\pi^*$ always picks $e_2$ and $e_3$ to achieve a value of $Q$. Hence, the worst-case cost  of $\pi^g$ is $\epsilon_a$, whereas the optimal solution has a cost of $2\epsilon_b$. Hence, $c_{wc}(\pi^g)=\frac{\epsilon_a}{2\epsilon_b} c^*$; one can select $\epsilon_a$ and $\epsilon_b$ to make $\frac{\epsilon_a}{\epsilon_b}$ arbitrarily large.

\section{Worst-Case Maximization Problem}
In this section, we study  a dual problem of the worst-case cover problem. We call this problem the worst-case adaptive submodular maximization problem. Our goal is to find a policy $\pi$ to maximize the worst-case utility $f_{wc}(\pi)$ subject to a budget constraint $B$, that is,
\[\max_{\pi: c_{wc}(\pi)\leq B} f_{wc}(\pi).\]

It is worth noting that the classic problem of maximizing a monotone submodular function subject to a budget constraint \citep{khuller1999budgeted} is a special case of our problem.

\begin{algorithm}[hptb]
\caption{Worst-Case Density-Greedy Policy $\pi^g$}
\label{alg:LPP4}
\begin{algorithmic}[1]
\STATE $t=1; \psi_0=\emptyset$.
\WHILE {true}
\STATE let $e_t \in \argmax_{e\in E} \frac{\Delta_{wc}(e\mid \psi_{t-1})}{c(e)}$;
\STATE $B=B-c(e_t)$;
\IF {$B\geq0$}
\STATE select $e_t$ and observe $\Phi(e_t)$;
\STATE update $\psi_{t}\leftarrow \psi_{t-1}\cup\{(e_t, \Phi(e_t))\}$; \label{line:1}
\ELSE
\STATE break; \COMMENT{Replace this line using ``select $e_t$; break;'' in the relaxed greedy policy $\pi^{g+}$.} \label{line:2}
\ENDIF
\STATE $t\leftarrow t+1$;
\ENDWHILE
\end{algorithmic}
\end{algorithm}

Our solution involves two candidate policies: one is a density-greedy based policy (labeled as $\pi^g$ by abuse of notation) and the other one selects a best singleton $e^{''}$ (i.e., $e^{''}\in \arg\max_{e\in E} \Delta_{wc}(e\mid \emptyset)$).   Our final algorithm (labeled as $\pi^a$) adopts the better one between these two candidates. Hence, the worst-case utility  of $\pi^a$ is $f_{wc}(\pi^a)=\max\{f_{wc}(\pi^g), \Delta_{wc}(e^{''}\mid \emptyset)\}$. To complete the design of $\pi^a$, we next explain $\pi^g$ in detail.

\textbf{Design of $\pi^g$.} Starting with round $t=1$ and observation $\psi_0=\emptyset$.
In each subsequent round $t$, $\pi^g$ selects an item $e_t$ that has the largest ``benefit-to-cost'' ratio, i.e.,
\[e_t \in \arg\max_{e\in E} \frac{\Delta_{wc}(e\mid \psi_{t-1})}{c(e)}.\] Next, we update the observation using $\psi_{t}\leftarrow \psi_{t-1}\cup\{(e_t, \Phi(e_t))\}$. This process iterates until  the budget constraint is violated. A detailed description of $\pi^g$ is listed in Algorithm \ref{alg:LPP4}.

For the purpose of proof, we  introduce a \emph{relaxed} version of $\pi^g$  (labeled as $\pi^{g+}$). $\pi^{g+}$ is identical to $\pi^g$ except that $\pi^{g+}$ allows to keep the first item that violates the budget constraint. Please refer to our comments added to Line \ref{line:2} in Algorithm \ref{alg:LPP4} for a detailed description of this difference.

We next analyze the performance of $\pi^a$. Before presenting the main theorem, we first provide a technical result.
\begin{theorem}
\label{thm:2-1}
If the utility function $f$  is worst-case monotone and worst-case submodular with respect to $p(\phi)$, then for any $\phi$ and any $L\leq c(E(\pi^{g+}, \phi))$,
 \begin{eqnarray}
\label{eq:last111}
h(\pi^{g+}_L\mid \phi)\geq (1-e^{-L/B})  f_{wc}(\pi^*),
\end{eqnarray} where $\pi^*$ is the optimal policy.
\end{theorem}
\emph{Proof:} We first recall some notations. Let $t[l]$ be the number of items that have a positive probability of being selected by $\pi^{g+}_l$ conditioned on $\phi$. Let $\psi_{t[l]-1}$ denote the partial realization of the first $t[l]-1$ items selected by $\pi^{g+}$ conditioned on $\phi$. To prove this theorem, it suffices to show that for all $l\in[L]$,
\begin{eqnarray}h(\pi^{g+}_l\mid \phi) - h(\pi^{g+}_{l-1}\mid \phi) \geq \frac{  f_{wc}(\pi^*)-f(\psi_{t[l]-1})}{B}. \label{eq:f1}
\end{eqnarray}

This is because by induction on $l$, we have that for any $L\leq c(E(\pi^{g+}, \phi))$,
\begin{eqnarray}
&&h(\pi^{g+}_L\mid \phi) = \sum_{l\in [L]} \left( h(\pi^{g+}_l\mid \phi) - h(\pi^{g+}_{l-1}\mid \phi) \right)~\nonumber\\
&&\geq (1-e^{-L/B})  f_{wc}(\pi^*). ~\nonumber
\end{eqnarray}

 Given $\psi_{t[l]-1}$, we adopt the same method as outlined in the proof of Theorem  \ref{thm:2} to construct $\phi^*$. Assuming $\pi^*$ selects $k$ items conditioned on $\phi^*$ such that $e^*_i$ represents the $i$-th item selected by $\pi^*$ conditioned on $\phi^*$. The following chain proves (\ref{eq:f1})
\begin{eqnarray*}
&&h(\pi^{g+}_l\mid \phi) - h(\pi^{g+}_{l-1}\mid \phi)\geq \frac{f_{wc}(\pi^*)-f(\psi_{t[l]-1})}{\sum_{i\in [k]}c(e^*_i)}\\
&& \geq \frac{  f_{wc}(\pi^*)-f(\psi_{t[l]-1})}{B},
\end{eqnarray*}
where the first inequality is derived from a similar proof as (\ref{eq:f}), with the only difference being that $Q$ is replaced with $f_{wc}(\pi^*)$; the second inequality is because the worst-case cost of $\pi^*$ is no larger than $B$. $\Box$

By the definition of $\pi^{g+}$, it always uses up the budget. This, together with the assumption that $f$ is minimal dependent, implies that $f_{wc}(\pi^{g+})\geq h(\pi^{g+}_B\mid \phi')$  where $\phi'$ is the worst-case realization of $\pi^{g+}$, i.e., $\phi'=\argmin_{\phi\in U^+} f(E(\pi^{g+}, \phi), \phi)$. This, in combination with Theorem \ref{thm:2-1}, leads to Corollary \ref{cor:2}.
\begin{corollary}
\label{cor:2}
If the utility function $f$  is worst-case monotone, worst-case submodular with respect to $p(\phi)$ and it satisfies the property of minimal dependency, then
 \begin{eqnarray}
\label{eq:last1131}
f_{wc}(\pi^{g+})\geq (1-e^{-1}) f_{wc}(\pi^*).
\end{eqnarray}
\end{corollary}

We next present the main theorem of this section.
\begin{theorem}
\label{thm:22}
If the utility function $f$  is worst-case monotone, worst-case submodular with respect to $p(\phi)$ and it satisfies the property of minimal dependency, then $f_{wc}(\pi^a) \geq \frac{1-e^{-1}}{2}f_{wc}(\pi^*)$.
\end{theorem}
\emph{Proof:} By the design of $\pi^a$, to prove this theorem, it suffices to show that  $\max\{f_{wc}(\pi^g), \Delta_{wc}(e^{''}\mid \emptyset)\}\geq \frac{1-e^{-1}}{2} f_{wc}(\pi^*)$.   Suppose $\phi^{''}$ is the worst-case realization of $\pi^g$, that is, $\phi^{''}=\argmin_{\phi} f(E(\pi^{g}, \phi), \phi)$. Let $\psi^{''}$ denote the partial realization of $E(\pi^{g}, \phi^{''})$ conditioned on $\phi^{''}$. Hence, $E(\pi^{g}, \phi^{''}) = \mathrm{dom}(\psi^{''})$ and $\phi^{''}\sim \psi^{''}$. 
Assume $e^{''}$ is the last item selected by the relaxed greedy policy $\pi^{g+}$ after observing $\psi^{''}$, that is, $E(\pi^{g+}, \phi^{''}) = E(\pi^{g}, \phi^{''})\cup \{e^{''}\}$. Let $\phi^\star(e^{''}) $ be the least favorable state for $e^{''}$ conditioned on $\psi^{''}$, i.e.,
\begin{eqnarray}
\phi^\star(e^{''}) = \argmin_{o\in O(e^{''}, \psi^{''})}f(\psi^{''} \cup \{(e^{''}, o)\}).
 \end{eqnarray}

By the definition of $f_{wc}(\pi^{g+})$ and the assumption that $f$ is minimal dependent, we have
\begin{eqnarray}
\label{Eq:wangsheng}
f\left(\psi^{''} \cup \{(e^{''}, \phi^\star(e^{''}))\}\right) \geq f_{wc}(\pi^{g+}).
 \end{eqnarray}

By the definition of $\phi^\star(e^{''})$, we have
 \begin{eqnarray}
f\left(\psi^{''} \cup \{(e^{''}, \phi^\star(e^{''}))\}\right) = f(\psi^{''}) + \Delta_{wc}(e^{''}\mid \psi^{''}).
 \end{eqnarray}

 It follows that
  \begin{eqnarray}
&&f\left(\psi^{''} \cup \{(e^{''}, \phi^\star(e^{''}))\}\right) = f(\psi^{''}) + \Delta_{wc}(e^{''}\mid \psi^{''})~\nonumber\\
&&=f_{wc}(\pi^g) + \Delta_{wc}(e^{''}\mid \psi^{''})~\nonumber\\
&&\leq f_{wc}(\pi^g) + \Delta_{wc}(e^{''}\mid \emptyset),\label{eq:wangsheng2}
 \end{eqnarray}
 where the second equality is due to $\phi^{''}$ being the worst-case realization of $\pi^g$ and the assumption that $f$ is minimal dependent; the inequality is due to $f$  being worst-case submodular with respect to $p(\phi)$ and the fact that $\emptyset\subseteq \psi^{''}$.

 Inequalities (\ref{Eq:wangsheng}) and (\ref{eq:wangsheng2}) jointly imply that
 \begin{eqnarray}
\label{Eq:wangsheng3}
f_{wc}(\pi^g) + \Delta_{wc}(e^{''}\mid \emptyset) \geq f_{wc}(\pi^{g+}).
 \end{eqnarray}

 This, together with Corollary \ref{cor:2}, implies that $f_{wc}(\pi^g) + \Delta_{wc}(e^{''}\mid \emptyset)  \geq f_{wc}(\pi^{g+}) \geq (1-e^{-1}) f_{wc}(\pi^*)$. Hence,
 \[\max\{f_{wc}(\pi^g), \Delta_{wc}(e^{''}\mid \emptyset)\}\geq \frac{1-e^{-1}}{2} f_{wc}(\pi^*).\] $\Box$

Note that the classic problem of maximizing a monotone submodular function subject to a budget constraint \citep{khuller1999budgeted} is a special case of our problem. The best approximation ratio for that problem, and therefore for ours, is $1-1/e$.
\section{Performance Evaluation}

\begin{figure*}[hptb]
\includegraphics[scale=0.26]{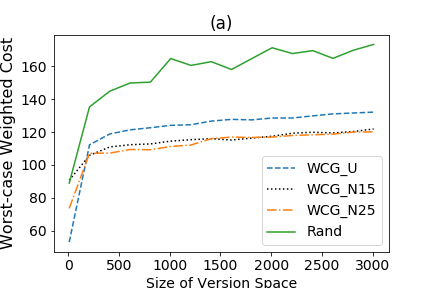}
\includegraphics[scale=0.26]{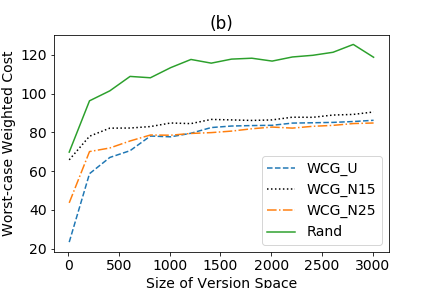}
\includegraphics[scale=0.26]{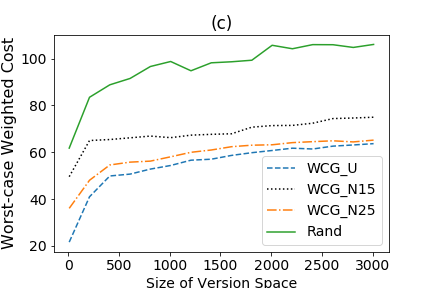}
\includegraphics[scale=0.26]{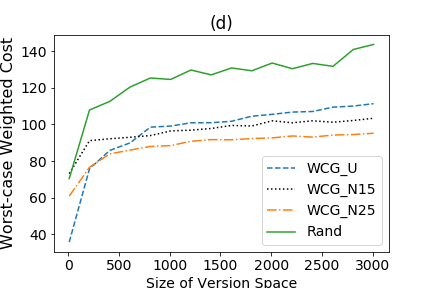}
\caption{Worst-case weighted cost vs. size of the version space $\mathcal{H}$}
\label{fig:cost}
\end{figure*}

\begin{figure*}[hptb]
\includegraphics[scale=0.26]{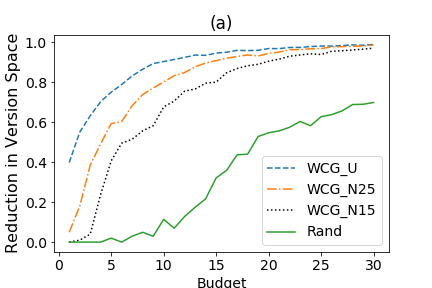}
\includegraphics[scale=0.26]{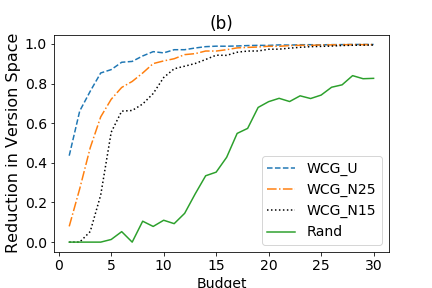}
\includegraphics[scale=0.26]{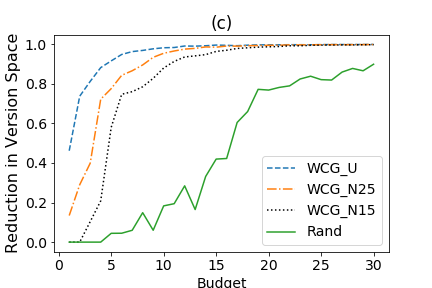}
\includegraphics[scale=0.26]{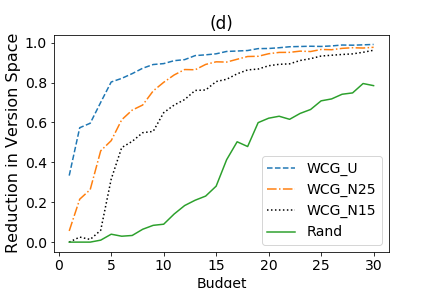}
\caption{Reduction in version space vs. budget $B$}
\label{fig:reduction}
\end{figure*}
In this section, we conduct experiments to evaluate the performance of our proposed Worst-Case Greedy (\emph{WCG}) algorithms in the context of active learning. Suppose we have a set of hypotheses $\mathcal{H}$ and a set of unlabeled data points $E$, where each $e\in E$ is selected randomly from a distribution $D$. In pool-based active learning, in order to reduce the expense of acquiring labeled data from domain experts, we select a sequence of data points to be labeled iteratively until the labels of all unlabeled examples can be inferred from the obtained labels. The version space is defined as the set of hypotheses that are consistent with the observed labels, and the cost of labeling a data point $e$ is a fixed value $c(e)$. Intuitively our goal is to minimize the worst-case cost of reducing the probability mass of the version space until the target hypothesis $h^*$ is pinpointed. Reducing the version space is achieved by eliminating false hypotheses through stochastic queries. For example, query $e\in E$ eliminates all hypotheses that do not agree with $h^*$ at $e$. For the budgeted version, our objective is to minimize the probability mass of the version space within a specific budget constraint.

Our first set of experiments evaluate the performance of our algorithm as measured by the worst-case cost with respect to the changes in the size of the version space $\mathcal{H}$, as shown in Figure \ref{fig:cost}. Each data point $e$ is assigned a value chosen randomly from its set of possible labels. The worst-case cost is calculated as the largest cost of pinpointing the target hypothesis $h^*$ after querying a sequence of data points. We consider three cost settings in our experiments. For the first setting, $c(e)$ is drawn from $(1,20)$ uniformly at random. The result is shown in the figure with label \emph{WCG\_U}. For the other two settings, $c(e)$ is drawn from $N(\mu,\sigma^2)$ with $\mu=7, \sigma=1.5$ and $\mu=7, \sigma=2.5$, respectively. Corresponding results are labeled as \emph{WCG\_N15} and \emph{WCG\_N25}, respectively, in the figure. To implement our algorithm, in each round we select a query with the largest conditional marginal utility over the cost until the target hypothesis is pinpointed. The conditional marginal utility is determined by  the worst-case reduction in version space, given the labels from past queries. A random algorithm is used as our baseline, which outputs a random sequence of queries until the target is pinpointed. For every set of experiments, we perform the simulation for 1,000 iterations and report the average results.

As shown in Figure \ref{fig:cost}, the $x$-axis refers to the size of the version space, ranging from $10$ to $3000$. The $y$-axis refers to the worst-case cost yielded by the corresponding algorithms. We evaluate our algorithm by using $20$ unlabeled data points and by varying the size of the label set. Figure \ref{fig:cost}(a) shows the results where each data point has binary labels. We observe that \emph{WCG} significantly outperforms the baseline in all test cases, yields a cost reduction of $30\%$ for binary labels. Note that our algorithm considers the marginal utility as well as the cost associated with each query, leading to a lower worst-case cost of the output sequence. We also observe that for smaller version space, on average our algorithm identifies the target hypothesis with fewer queries, and \emph{WCG\_U} benefits from taking more low-cost queries since the algorithm prefers a larger marginal utility to cost ratio. As the size of the version space increases, however, \emph{WCG\_U} yields a much higher cost as taking low-cost queries alone is not enough to pinpoint the target and queries with potentially high cost are required to further reduce the version space.

We observe a similar structure in Figure \ref{fig:cost}(b), (c) and (d), showing the results for three-label data points, four-label data points and a hybrid case, respectively. For the hybrid case, we randomly divide our $20$ unlabeled data points into three groups. The first group contains $10$ binary-label data points, the second group contains $5$ three-label data points, and the third group contains $5$ four-label data points. We observe that our algorithm generates a lower worst-case cost when each data point has more possible labels. The reason is that queries with more possible labels tend to yield a higher marginal reduction in version space, therefore less queries are selected in the output, leading to a lower worst-case cost.

Our second set of experiments investigate how the budget affects the reduction in version space, as illustrated in Figure \ref{fig:reduction}. The $x$-axis holds the value of the budget, and the $y$-axis holds the reduction in version space generated by the algorithms. We consider $3000$ hypothesis with $20$ unlabeled data points, and tight budget constraint is enforced. Figure \ref{fig:reduction}(a), (b), (c) and (d) plot the results for binary-label data points, three-label data points, four-label data points and the hybrid case as aforementioned, respectively. As anticipated, the reduction in version space becomes greater as the budget increases. Again, for smaller budgets, \emph{WCG} yields a higher reduction in version space under uniform cost model than it does under the other two cost models. As the budget goes up, more queries are included in the output sequence, and we observe that the reduction in version space among different cost models converges.

\section{Appendix}
Proof of  Lemma \ref{lem:end}.   For the case when $c_1 \leq c_3$, this result has been proved in Lemma 2 in \cite{tang2016optimizing}. We next focus on the case when $c_1 > c_3$. We prove this lemma in five subcases depending on relation between $x$ and the other four constants. Notice that when $c_1 > c_3$, $c_1\geq c_2$, $c_3 \geq c_4$, and $c_2 \leq c_4$, we have $c_1 > c_3 \geq c_4 \geq c_2$.
\begin{itemize}
\item If $x \geq c_1 > c_3 \geq c_4 \geq c_2$, then $\min\{c_1, x\}=c_1$, $\min\{c_2, x\}=c_2$, $\min\{c_3, x\}=c_3$ and $\min\{c_4, x\}=c_4$. Thus, $\min\{c_1, x\}-\min\{c_2, x\}\geq\min\{c_3, x\}-\min\{c_4, x\}$ due to the assumption that $c_1-c_2\geq c_3-c_4$.
\item If $c_1 > x \geq c_3 \geq c_4 \geq c_2$, then $\min\{c_1, x\}=x$, $\min\{c_2, x\}=c_2$, $\min\{c_3, x\}=c_3$ and $\min\{c_4, x\}=c_4$. Thus, $\min\{c_1, x\}-\min\{c_2, x\} = x- c_2$ and $\min\{c_3, x\}-\min\{c_4, x\}=c_3-c_4$. Because $x\geq c_3$ and $c_2\leq c_4$, we have $x- c_2 \geq c_3-c_4$. It follows that $\min\{c_1, x\}-\min\{c_2, x\}\geq\min\{c_3, x\}-\min\{c_4, x\}$.
\item If $c_1 >c_3 > x \geq  c_4 \geq c_2$, then $\min\{c_1, x\}=x$, $\min\{c_2, x\}=c_2$, $\min\{c_3, x\}=x$ and $\min\{c_4, x\}=c_4$. Thus, $\min\{c_1, x\}-\min\{c_2, x\} = x- c_2$ and $\min\{c_3, x\}-\min\{c_4, x\}=x-c_4$. Because $c_2 \leq c_4$, we have $x- c_2 \geq x-c_4$, thus, $\min\{c_1, x\}-\min\{c_2, x\}\geq\min\{c_3, x\}-\min\{c_4, x\}$.
\item If $c_1 >c_3 \geq  c_4 > x  \geq c_2$, then $\min\{c_1, x\}=x$, $\min\{c_2, x\}=c_2$, $\min\{c_3, x\}=x$ and $\min\{c_4, x\}=x$. Thus, $\min\{c_1, x\}-\min\{c_2, x\} = x- c_2$ and $\min\{c_3, x\}-\min\{c_4, x\}=x-x =0$. Because $x  \geq c_2$, we have $x- c_2 \geq 0$, thus, $\min\{c_1, x\}-\min\{c_2, x\}\geq\min\{c_3, x\}-\min\{c_4, x\}$.
\item If $c_1 >c_3 \geq  c_4 \geq c_2 > x  $, then $\min\{c_1, x\}=x$, $\min\{c_2, x\}=x$, $\min\{c_3, x\}=x$ and $\min\{c_4, x\}=x$. Thus, $\min\{c_1, x\}-\min\{c_2, x\} = x- x = 0$ and $\min\{c_3, x\}-\min\{c_4, x\}=x-x =0$. Thus, $\min\{c_1, x\}-\min\{c_2, x\}\geq\min\{c_3, x\}-\min\{c_4, x\}$.
\end{itemize}
\bibliographystyle{ijocv081}
\bibliography{reference}
\clearpage

\end{document}